\documentclass[reprint,aps,prl,superscriptaddress,amsmath,amssymb,floatfix,footinbib,longbibliography]{revtex4-1}
\usepackage[bottom]{footmisc} 
\usepackage{amssymb,bbold,dsfont}
\usepackage{amsmath}
\usepackage{color}
\usepackage{graphicx}
\usepackage{marvosym}
\usepackage{ulem}
\usepackage{soul}
\usepackage{notoccite}
\usepackage{comment}
\usepackage{pifont}
\usepackage{bbm}
\usepackage{graphicx,tabularx}
\usepackage{dcolumn}
\usepackage{bm}
\usepackage{xcolor}
\usepackage{marvosym}
\usepackage[breaklinks=true,colorlinks,citecolor=blue,linkcolor=blue,urlcolor=blue]{hyperref}
\usepackage{titlesec}
\newcommand{\be}{\begin{equation}}
\newcommand{\ee}{\end{equation}}

\renewcommand{\vec}[1]{{\bf #1}}
\renewcommand{\bm}[1]{{\bf #1}}
\def \bea{\begin{eqnarray}}
\def \eea{\end{eqnarray}}

\begin{document}
\title{Nonlinear bulk photocurrent probe ${\cal Z}_2$ topological phase transition}
        \author{Debasis Dutta}
	  \email{debasis.dutta@ntu.edu.sg}
        \affiliation{Department of Physics, Indian Institute of Technology  Kanpur, Kanpur-208016, India}
        \thanks{D. D. and R. A. contributed equally to this project.}
 \affiliation{Division of Physics and Applied Physics, School of Physical and Mathematical Sciences,
 Nanyang Technological University, Singapore 637371}
        \author{Raihan Ahammed}
	\email{raihan@iitk.ac.in}
    \affiliation{Department of Physics, Indian Institute of Technology Kanpur, Kanpur-208016, India}
\author{Yingdong Wei} 
\affiliation{State Key Laboratory of Infrared Physics, Shanghai Institute of Technical Physics, Chinese Academy of Sciences, 500 Yu-Tian Road, Shanghai 200083, China.}
\author{Xiaokai Pan}
\affiliation{State Key Laboratory of Infrared Physics, Shanghai Institute of Technical Physics, Chinese Academy of Sciences, 500 Yu-Tian Road, Shanghai 200083, China.}
\author{Xiaoshuang Chen}
\affiliation{State Key Laboratory of Infrared Physics, Shanghai Institute of Technical Physics, Chinese Academy of Sciences, 500 Yu-Tian Road, Shanghai 200083, China.}
\author{Lin Wang} 
\email{wanglin@mail.sitp.ac.cn} 
\affiliation{State Key Laboratory of Infrared Physics, Shanghai Institute of Technical Physics, Chinese Academy of Sciences, 500 Yu-Tian Road, Shanghai 200083, China.}

\author{Amit Agarwal}
\email{amitag@iitk.ac.in}
\affiliation{Department of Physics, Indian Institute of Technology Kanpur, Kanpur-208016, India}
\date{\today}
\begin{abstract} 
Detecting topological phase transitions in bulk is challenging due to the limitations of surface-sensitive probes like ARPES. Here, we demonstrate that nonlinear bulk photocurrents, specifically shift and injection currents, serve as effective probes of ${\cal Z}_2$ topological transitions. These photocurrents show a robust polarity reversal across the ${\cal Z}_2$ phase transition, offering a direct optical signature that distinguishes strong topological phases from weak or trivial ones. This effect originates from a reorganization of key band geometric quantities, the Berry curvature and shift vector, on time-reversal-invariant momentum planes. Using a low-energy Dirac model, we trace this behaviour to a band inversion in the time-reversal-invariant momentum plane that drives the topological transition. We validate these findings through tight-binding model for Bi$_2$Te$_3$ and first-principles calculations for ZrTe$_5$ and BiTeI, where the topological phase can be tuned by pressure or temperature. Our results establish nonlinear photocurrent as a sensitive and broadly applicable probe of ${\cal Z}_2$ topological phase transitions.
%
\end{abstract}
\maketitle

\begin{figure}[t!]
    \centering    
    \includegraphics[width=0.92\linewidth]{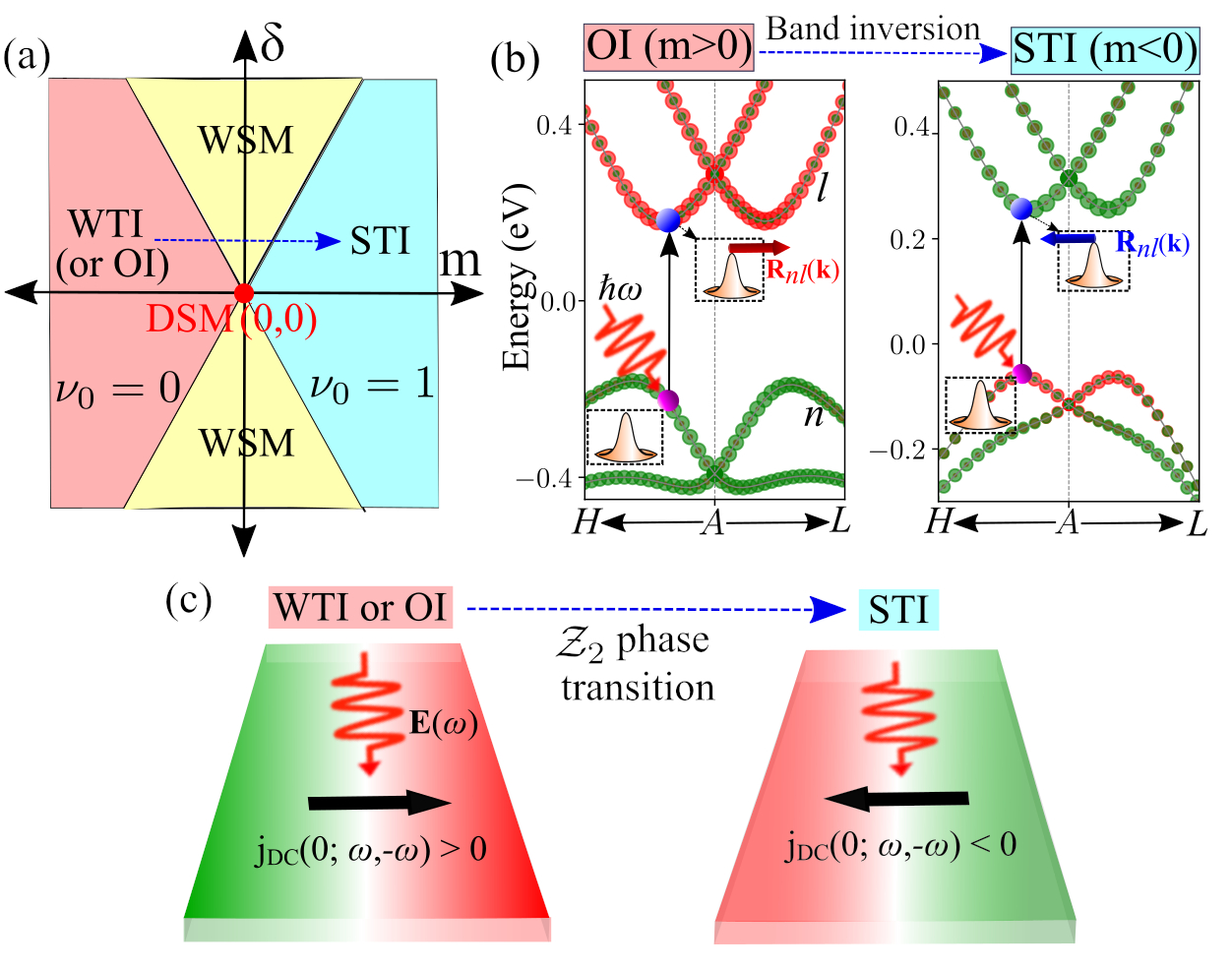}
    \caption{
    \textbf{Signature of topological phase transition.} 
(a) Phase diagram of the Z$_2$ topological transition as a function of the bandgap parameter $m$ and the inversion symmetry-breaking strength $\delta$, highlighting distinct phases including Dirac semimetal (DSM), Weyl semimetal (WSM), STI, and WTI. 
(b) Band inversion mechanism in BiTeI, accompanied by reversal of the bandgap sign between the ordinary ($\nu_0 = 0$) and strong topological ($\nu_0 = 1$) phases. 
Orbital projections of Bi-$p_z$ (red) and Te/I-$p_z$ (green) are indicated. The Berry curvature and the shift vector undergo significant reorganization across the ${\cal Z}_2$ transition. (c) This band geometric reorganization drives the polarity reversal in the bulk photovoltaic current across the ${\cal Z}_2$ phase transition.}
\label{Figure1} 
\end{figure}

\textcolor{blue}{\it Introduction.-} 
Topological phase transitions (TPTs) represent a distinct class of quantum critical phenomena beyond the conventional Landau-Ginzburg paradigm. They are characterized by abrupt changes in global topological invariants encoded in the band geometry of electronic wavefunctions. In 3D time-reversal symmetric insulators, such transitions are classified by four ${\cal Z}_2$ indices $(\nu_0; \nu_1, \nu_2, \nu_3)$, distinguishing strong topological insulators (STIs), weak topological insulators (WTIs), and ordinary insulators (OIs)~\cite{Fu_Kane_PRL2007, LiangFu2007T, HasanRMP2010, Vanderbilt2011T}. Among these, the strong index $\nu_0$ is robust against time-reversal-invariant perturbations and plays a key role in protecting surface states. These transitions typically involve a bulk band inversion driven by a tunable system parameter [see Fig.~\ref{Figure1}(a)]. They can be induced by tuning temperature, pressure, doping, electric field, or strain~\cite{Murakami2017, BiTeI_TPT_2013, Liu2014, Fan2017,Shaffique2018E,Monserrat2019,Cheng2019Science,PhysRevLett.110.176401,Singh_PRL_2018}.

Detecting TPTs in bulk materials, however, poses significant experimental challenges~\cite{Xu-yangXu2011, PhysRevB.100.201107}. Conventional probes such as angle-resolved and spin-resolved photoemission are surface-sensitive and often limited by sample quality and cleaving conditions~\cite{Wang_2024}. This has led to growing interest in alternative approaches such as high-harmonic generation~\cite{Heide2022} and nonlinear transport~\cite{Sinha_B_2022, atasi2022N, Adak2024, Surat2025Q}. These are inherently bulk-sensitive and directly couple to the underlying topological band structure and symmetry of electronic states.

\begin{figure*}[t!]
    \centering    
    \includegraphics[width=0.9\linewidth]{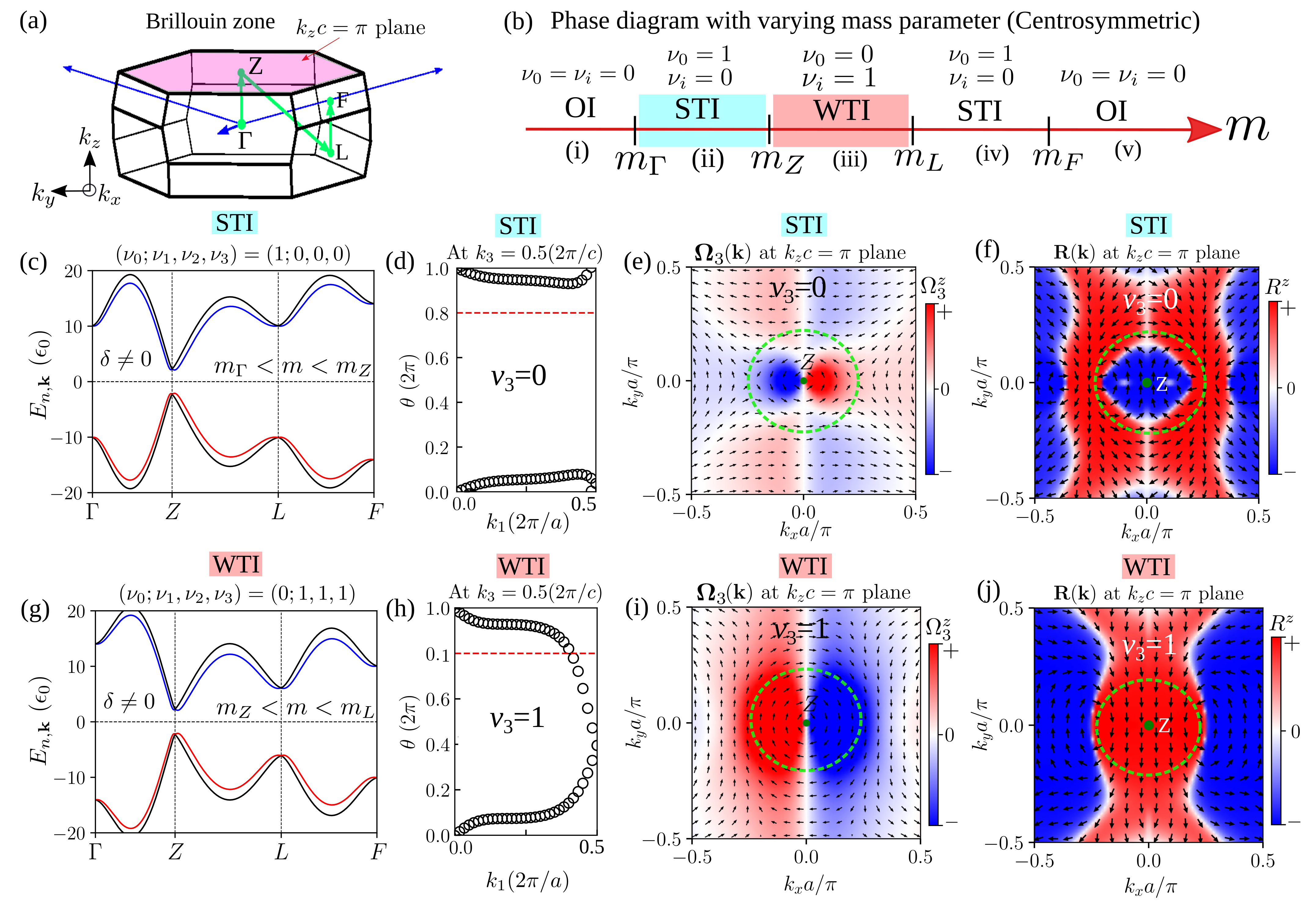}
    \caption{
\textbf{Reorganization of band geometry across the ${\cal Z}_2$ transition in Bi$_2$Te$_3$.} 
(a) Rhombohedral Brillouin zone for a Bi$_2$Te$_3$-like system, showing the time-reversal invariant momentum (TRIM) points. 
(b) Phase diagram for inversion-symmetric Bi$_2$Te$_3$, with transitions at critical values of the mass parameter $m = m_i$, corresponding to band inversions at different TRIM points $i \in \{\Gamma, Z, F, L\}$. 
(c--d) Band structure and hybrid Wannier center evolution, without inversion symmetry, in the strong topological phase with ${\cal Z}_2 = (1;0,0,0)$, confirming $\nu_3 = 0$ in the regime $m_{\Gamma} < m < m_Z$ (cyan region in panel b). Here, {$\epsilon_0$ represents hopping amplitude} and $a,~ c$ denote lattice parameters of the system. 
(e--f) Berry curvature $\bm{\Omega}_3(\bm{k})$ of the first conduction band and total shift vector $\vec{R}(\vec{k})$ for $x$-polarized light, evaluated on the $k_z = \pi/c$ plane in the strong topological phase. The in-plane components are shown as arrows, and the out-of-plane components are shown as a colormap. 
(g--j) Corresponding results for the weak topological phase [${\cal Z}_2 = (0;1,1,1)$] with $\nu_3 = 1$, showing sharp reorganization of both band geometric quantities in the TRIM plane hosting the band inversion at $Z$ point (highlighted with green dashed circles).
}
\label{Figure2}
\end{figure*} 

Here, we demonstrate that the bulk photovoltaic effect (BPE), manifesting as nonlinear DC photocurrents in noncentrosymmetric materials, offers a powerful probe of ${\cal Z}_2$ band topology. Prominent contributions to these photocurrents are shift and injection currents, which originate from band geometric quantities such as the Berry curvature and shift vector~\cite{Sipe2000S, AhnPRX2020, Xiong_2021, Agarwal2022A, Dai2023, Ma2023, FrankKoppen2025T}. We show that BPEs exhibit a robust polarity reversal across ${\cal Z}_2$ topological transitions between STI and WTI or OI phases. This reversal is driven by a reorganization of the Berry curvature and shift vector in the time-reversal-invariant momentum (TRIM) planes associated with band inversion accompanying the phase transition [Fig.~\ref{Figure1}(b–c)]. We validate this mechanism using tight-binding analysis on Bi$_2$Te$_3$ and first-principles calculations on ZrTe$_5$ which undergo pressure and temperature-tunable TPTs~\cite{Monserrat2019, TPT_ZrTe5_exp_PRL_paper, Zhang2017E, PhysRevLett.126.016401}, and in bulk BiTeI under volume expansion~\cite{Bahramy2012}. Our results establish a nonlinear optical signature for detecting and utilizing ${\cal Z}_2$ transitions in quantum materials with potential applications in tunable terahertz photodetectors~\cite{LinWang2021, Libo2021, DANG20222659}.


\textcolor{blue}{\it Shift and injection currents.–} 
When monochromatic light with electric field $\vec{E}(t) = \vec{E}(\omega)e^{-i\omega t} + \vec{E}^{*}(\omega)e^{i\omega t}$ illuminates a noncentrosymmetric crystal, it generates a second-order DC photocurrent~\cite{Sipe2000S}. This photocurrent arises from interband transitions and is expressed as 
\begin{equation}
    j^a_{\text{DC}} = \sigma_{abc}(0; \omega, -\omega) E_b(\omega) E_c^{*}(\omega),
\end{equation}
where $\sigma_{abc}$ is the third-rank nonlinear conductivity tensor.
In time-reversal symmetric systems, the dominant contributions to BPE are the shift and injection currents, which manifest under linearly and circularly polarized light, respectively~\cite{Sipe2000S, Ivosuza2018, Grushin2017, AhnPRX2020, Wanatabe_PhysRevX2021, Agarwal2022A, Ma2023}. These arise from shift in position (shift current) and change in velocity (injection current) of electrons during interband optical transitions.

The corresponding conductivities are expressed as
\[
\sigma^{\text{shift}/\text{inj}}_{abc} = \beta_0 \int_{\vec{k}} \tilde{\sigma}^{\text{shift}/\text{inj}}_{abc}, \quad \text{with } \beta_0 = \frac{\pi e^3}{\hbar^2}~,
\]
with Brillouin zone integral defined as $\int_{\vec{k}} \equiv \int d^d k/(2\pi)^d$  in $d$ dimensions. The $\vec{k}$-resolved integrands are 
\begin{eqnarray}
\tilde{\sigma}_{abc}^{\text{shift}} &=& - \sum_{n \ne m} f_{nm} \left( R^{a;c}_{nm} + R^{a;b}_{nm} \right) r^c_{nm} r^b_{mn} \delta (\omega_{mn} - \omega), \nonumber \\
\tilde{\sigma}_{abc}^{\text{inj}} &=& -2\tau \sum_{n \ne m} f_{nm} \Delta^a_{nm} r^c_{nm} r^b_{mn} \delta (\omega_{mn} - \omega)~. \label{Eq_inj}
\end{eqnarray}
Here, $\omega_{mn} = (E_m - E_n)/\hbar$, $f_{nm} = f_n - f_m$ is the difference in band occupations, and $r^b_{nm} = -iv^b_{nm}/\omega_{mn}$ is the dipole matrix element, with $v^b_{nm} = \hbar^{-1} \langle u_n | \partial_{k_b} H(\vec{k}) | u_m \rangle$, and $\tau$ denotes the scattering time.

The shift current is governed by the real-space shift vector~\cite{Sipe2000S, AhnPRX2020, Xiong_2021}
\begin{equation}
    \vec{R}_{nm} = \vec{A}_{nn} - \vec{A}_{mm} + i \nabla_{\vec{k}} \log r_{nm}~,
\end{equation}
where $\vec{A}_{nn} = i \langle u_n | \nabla_{\vec{k}} | u_n \rangle$ is the Berry connection of the $n'th$ band and $r_{nm}$ is the dipole amplitude projected along the light polarization direction  $\hat{\vec{e}}$.

In contrast, the injection current depends on the velocity difference $\Delta^a_{nm} = v^a_{nn} - v^a_{mm}$ and the Berry curvature,
\begin{equation}
    \Omega^a_n = -2 \epsilon_{abc} \, \mathrm{Im} \sum_{m \ne n} \langle \partial_{k_b} u_n | u_m \rangle \langle u_m | \partial_{k_c} u_n \rangle~.
\end{equation}
In time-reversal symmetric systems lacking inversion symmetry, $\sigma^{\text{shift}}_{abc}$ is real, while $\sigma^{\text{inj}}_{abc}$ is purely imaginary~\cite{deJuan2017, Ma2023,AhnPRX2020, Wanatabe_PhysRevX2021}. Importantly, the shift vector and Berry curvature reorganize sharply across ${\cal Z}_2$ topological phase transitions, leading to observable sign reversals in nonlinear photocurrents.

\begin{figure}[t!]
    \centering    
    \includegraphics[width=\linewidth]{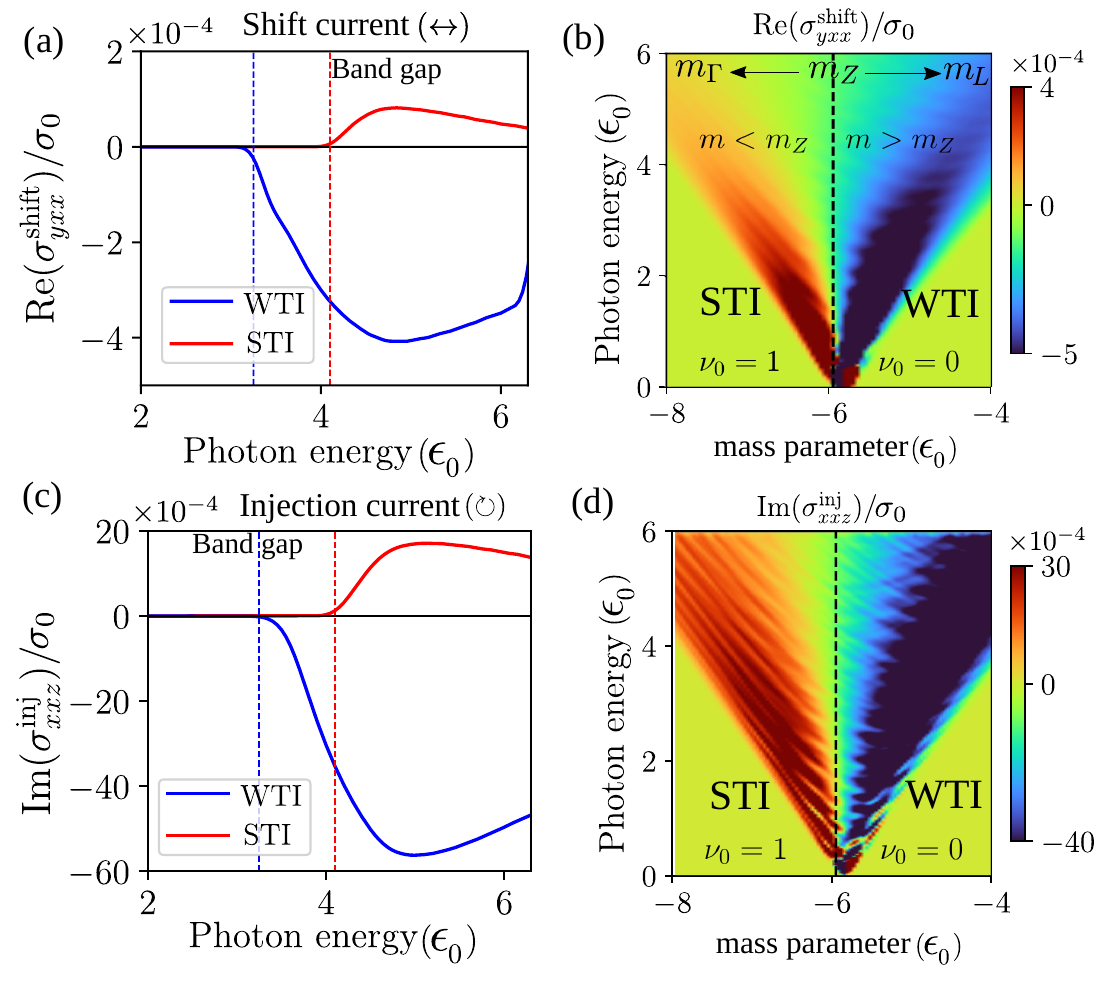}
    \caption{
    \textbf{Polarity reversal of photocurrent across the ${\cal Z}_2$ phase transition.} 
    (a) Calculated shift conductivity ${\rm Re}(\sigma_{yxx}^{\rm shift})$ as a function of photon energy for representative mass parameters $m$ in the strong ($m < m_Z$) and weak ($m > m_Z$) topological regimes of Bi$_2$Te$_3$. Here, $\sigma_0=\pi e^3/(\hbar\epsilon_0)$.
    (b) Colormap of ${\rm Re}(\sigma_{yxx}^{\rm shift})$ as a function of photon energy and mass parameter $m$, showing a distinct sign reversal between the strong ($\nu_0 = 1$) and weak ($\nu_0 = 0$) topological phases.  
    (c--d) Line and colormap plots of the injection conductivity ${\rm Im}(\sigma_{xxz}^{\rm inj})$ versus $m$ and photon energy, also exhibiting a topology-sensitive sign flip. These results demonstrate that both shift and injection conductivities undergo polarity reversals across the topological phase boundary, serving as a direct nonlinear optical signature of the ${\cal Z}_2$ topological phase transition.}
    \label{Figure_TB}
\end{figure}

\begin{figure*}[t!]
    \centering    
    \includegraphics[width=\linewidth]{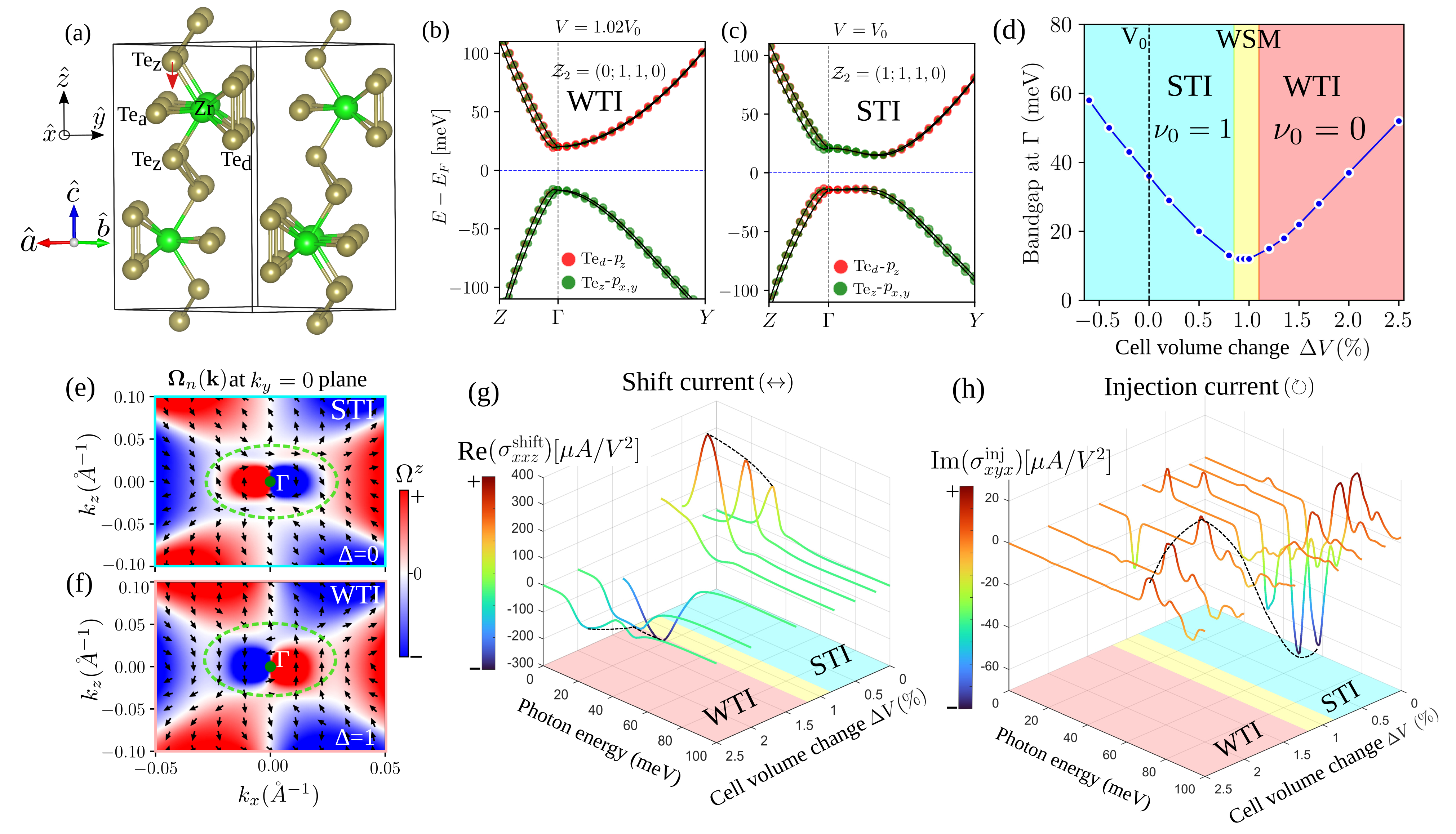}
    \caption{
    \textbf{Photocurrent reversal across the ${\cal Z}_2$ topological phase transition in ZrTe$_5$.} 
    (a) Primitive unit cell of ZrTe$_5$, with crystalline $c$ axis aligned with the $z$ axis. Inversion symmetry is broken by the displacement of Te$_z$ atoms~\cite{Ando2022}. 
    (b,c) Orbital-resolved band structures at relaxed and expanded volumes show a transition from weak to strong topology via band inversion at $\Gamma$ between Te$_d$-$p_z$ and Te$_z$-$p_{x,y}$ orbitals. The location of different Te atoms (Te$_d$ and Te$_z$) is indicated in (a).
    (d) Evolution of the bulk bandgap with volume change $\Delta V \equiv (V - V_0)/V_0$ indicates a tunable ${\cal Z}_2$ phase transition. 
    (e--f) Berry curvature {for the first valence band} in the $k_y = 0$ plane reverses orientation across the transition, reflecting a change in the 2D ${\cal Z}_2$ invariant $\Delta$ from 0 to 1. 
    (g--h) Shift and injection conductivities calculated from first principles, as functions of photon energy, exhibit clear polarity reversal in the two phases (see the black dashed line). This is driven by interband transitions at $\Gamma$, which is the center of the band inversion across the topological phase boundary.  These results establish a direct connection between band topology and nonlinear photocurrents in ZrTe$_5$.}
    \label{Figure_ZrTe5}
\end{figure*}

\textcolor{blue}{\it Topological phase transitions in Bi$_2$Te$_3$.-}  We now examine nonlinear photocurrent response in Bi$_2$Te$_3$ across the ${\cal Z}_2$ topological phase transition. For this, we employ a minimal tight-binding model that captures essential features of Bi$_2$Te$_3$-class materials with rhombohedral lattice symmetry~\cite{Zhang2009, KuramotoPRB2011}. We work with the hybridized atomic orbitals $P_z^{+/-}$ for the $(\uparrow,\downarrow)$ spins. Using this as a basis set,  $\left(\left|P_z^+,\uparrow\right\rangle,\left|P_z^-,\uparrow\right\rangle,\left|P_z^+,\downarrow\right\rangle,\left|P_z^-,\downarrow\right\rangle\right)$, the $4\times4$ Hamiltonian is given by, 
\be
H(\bm{k},m)=h_{0}+\sum_{i=1}^{5} h_{i}(\bm{k})\Gamma_{i} + m(\mathbb{1}\otimes \tau_{3})+\delta(\mathbb{1}\otimes{\tau_1} + \sigma_{1}\otimes\tau_{2})~.
\label{Eq_Ham_main_text}
\ee
Here, 
$m$ is a tunable mass parameter that controls the bandgap and drives the phase transition, while $\delta$ introduces inversion symmetry (IS) breaking.
The Pauli matrices $\sigma_i$ and $\tau_i$ act on spin and orbital spaces, and $\Gamma_i$ are mutually anticommuting Dirac matrices. Other details of the hopping terms $h_i(\bm{k})$ and the tight-binding model are provided in Sec.~S1 of the SM \footnote{Supplementary materials (SM) provide additional information on the following: i) Details of the tight binding model, ii) Computation of ${\cal Z}_2$ invariants from hybrid Wannier charge centers, iii) Details of density functional theory based calculations, iv) Numerical calculation of shift and injection current from Wannier function, v) Crystal symmetry constraint on the nonlinear conductivity tensor.
}. 

Tuning the mass parameter $m$ drives the system between OI, STI, and WTI phases. In the IS-preserving limit ($\delta = 0$), the bandgap closes at TRIM points ($\Gamma$, $Z$, $L$, $F$), marking transitions through Dirac semimetal phases. With broken IS ($\delta \ne 0$), the system transitions via a Weyl semimetal phase with gap closure shifted away from TRIM points~\cite{ShuichiMurakami_2007}. 
We classify topological phases using one strong index $\nu_0$ and three weak indices $(\nu_1, \nu_2, \nu_3)$, computed via the evolution of the hybrid Wannier charge centers (see Sec.~S3 of the SM~\cite{Note1} for details). For STI, $\nu_0 = 1$; for WTI, $\nu_0 = 0$, and at least one weak index is nonzero.

Figure~\ref{Figure2}(b) shows the phase diagram of the TB-model of Bi$_2$Te$_3$. At $m < m_{\Gamma}$, the system is an OI. For $m > m_{\Gamma}$, band inversion at $\Gamma$ yields a STI with ${\cal Z}_2=(1;0,0,0)$. Further increasing $m > m_Z$ leads to a second band inversion at $Z$, resulting in a WTI phase with ${\cal Z}_2=(0;1,1,1)$. The corresponding band dispersions are shown in Figs.~\ref{Figure2}(c) and (g). This transition closes and reopens the gap in the $k_z = \pi/c$ plane, flipping the weak index $\nu_3$ as shown in Figs.~\ref{Figure2}(d) and (h). Importantly, this flip manifests as a reversal in the Berry curvature and shift vector fields across the $Z$ point as highlighted in Figs.~\ref{Figure2}(e–j)\footnote{But, at $k_z=0$ plane, around $\Gamma$ point, the distribution of Berry curvature and shift vector fields are identical at both STI and WTI phase, where the 2D ${\cal Z}_2$ invariant (see EM1) $\Delta=1$ for both phase, as shown in Fig.~S2 of the SM~\cite{Note1}.}. 
Such reorganization of the BC distribution and the total shift vector in the TRIM plane ($k_z=\pi/c$ here) can also be intuitively understood from a simple low-energy model calculation presented in EM2 of the End Matter.

The total shift vector field, which reverses polarity across the phase transition in Figs.~\ref{Figure2}(f)-(j) is defined by summing over all bands involved in optical transitions, 
\be
\vec{R}(\vec{k}) = \sum_{m \ne n} f_m(1 - f_n) \vec{R}_{mn}(\vec{k})~.
\ee
Here, $f_m$ and $f_n$ denote the Fermi-Dirac occupation factor for bands  $m$ and $n$. The sign reversal in $\vec{R}(\vec{k})$ leads to a corresponding sign change in the DC shift photocurrent [Figs.~\ref{Figure_TB}(a-b)]. Likewise, the injection current ${\rm Im}(\sigma^{\rm inj}_{xxz})$ changes sign [Figs.~\ref{Figure_TB}(c-d)]. 
These behaviours highlight the intrinsic connection between band topology and quantum geometry-driven nonlinear photocurrents. 

\textcolor{blue}{\it Photocurrent reversal across transition in ZrTe$_5$.-} 
To demonstrate the generality of this mechanism beyond Bi$_2$Te$_3$, we now turn to ZrTe$_5$, a layered material with a pressure/temperature tunable ${\cal Z}_2$ topological phase transition~\cite{Monserrat2019, Cheng2019Science, Fan2017, Krausz2024, Zhang2017E}. ZrTe$_5$ is a layered van der Waals material with orthorhombic crystal structure (space group \textit{Cmcm}; lattice constants $a=3.979$~\AA, $b=14.470$~\AA, $c=13.676$~\AA)~\cite{FJELLVAG198691}, with layers stacked along the $b$-axis [Fig.~\ref{Figure_ZrTe5}(a)]. Although the \textit{Cmcm} structure preserves inversion symmetry, recent studies~\cite{Ando2022, Guoqing_ZrTe5_2023, PhysRevB.111.L041201} report symmetry breaking due to Te atomic displacements, lowering the symmetry to non-centrosymmetric \textit{Cm} group with only one mirror plane ($M_{yz}$). 

At the relaxed volume $V_0$, ZrTe$_5$ is a STI with topological indices ${\cal Z}_2 = (1;1,1,0)$ and a bulk bandgap $E_g = 36$ meV at $\Gamma$ [Fig.~\ref{Figure_ZrTe5}(c)]. As the unit-cell volume increases, the gap closes and reopens, 
driving a transition into a WTI phase with ${\cal Z}_2 = (0;1,1,0)$ [see Fig.~\ref{Figure_ZrTe5}(d) and Sec.~S4 of the SM]. This STI–WTI transition is driven by a band inversion near $\Gamma$ between Te$_d$-$p_z$ and Te$_z$-$p_{x,y}$ orbitals [Figs.~\ref{Figure_ZrTe5}(b–c)], and can be modeled by a Dirac-like Hamiltonian with mass parameter $m$, where $m<0$ ($m>0$) corresponds to the STI (WTI) phase [see EM2 for details].

This band inversion alters the 2D ${\cal Z}_2$ invariant ($\Delta$) on the $k_y = 0$ plane, changing from $\Delta = 0$ (STI) to $\Delta = 1$ (WTI), while keeping the invariant unchanged ($\Delta=1$) on the $k_y=\pi/b$ TRIM plane for both STI and WTI phases (see Fig.~S4 of SM~\cite{Note1}). Different values of $\Delta$ at $k_y=0$ and $k_y=\pi/b$ planes make $\nu_0=1$ in STI phase, while $\nu_0=0$ for WTI phase, having $\Delta=1$ at both planes (see EM1 for details). This change is also reflected in the Berry curvature field, which reverses orientation across the transition in the $k_y = 0$ plane where $\Delta$ changes [Figs.~\ref{Figure_ZrTe5}(e–f)], but remains unchanged at $k_y=\pi/b$ where $\Delta$ does not change (see Fig.~S5 of  SM~\cite{Note1}).

This reorganization of band geometry leads to a measurable polarity reversal in nonlinear optical responses. In Fig.~\ref{Figure_ZrTe5}(g), we show the linear shift conductivity ${\rm Re}(\sigma^{\rm shift}_{xxz})$ as a function of photon energy for various cell volumes across the transition. A clear polarity reversal is observed across the STI to WTI phase transition, coinciding with the change in strong index $\nu_0$ from $1$ to $0$. The circular injection conductivity ${\rm Im}(\sigma^{\rm inj}_{xyx})$ shows a similar sign flip near 40–60 meV [Fig.~\ref{Figure_ZrTe5}(h)], corresponding to interband transitions near $\Gamma$ where band inversion occurs (see Sec.~S4 and S5 of SM~\cite{Note1} for symmetry analysis and details of calculations). 

The magnitude of shift conductivities in Fig.~\ref{Figure_ZrTe5}(g) reaches about $ 300$ $\mu A/V^2$, comparable to the observed values of bulk photocurrent in 3D materials such as BaTiO$_3$ and GaAs~\cite{PhysRevLett.133.186801, Dai2023}. The calculated circular injection conductivities are also of similar magnitude and experimentally  accessible~\cite{Orenstein2020}. These results establish ZrTe$_5$ as a robust platform for experimentally probing TPT using terahertz light (photon energies between $0.3$-$40$ meV). The polarity-reversing photocurrent in ZrTe$_5$ could enable applications in terahertz photodetection~\cite{PhysRevLett.113.096601, THZ_Detection_exp1}, rectification~\cite{Libo2021}, and related optoelectronic technologies. 

We further validate this behavior in BiTeI, which offers an independent test case for an OI to STI transition. We show the sign flip in the photocurrent in BiTeI for the two phases in Fig.~\ref{Fig_BiTeI} in EM3.  

\textcolor{blue}{\it Conclusion.–} 
We have demonstrated that nonlinear photocurrents, arising from band geometric quantities such as the shift vector and Berry curvature, exhibit polarity reversals across ${\cal Z}_2$ topological phase transitions (see Table~1 in EM). This effect results from a sharp reorganization of these quantities on momentum planes where the band inversion associated with the phase transition occurs. The resulting nonlinear optical signature provides a powerful and symmetry-sensitive probe of topological order, complementary to transport experiments and photoemission spectroscopy. 

These findings open up new opportunities for broadband optical probes, such as terahertz emission spectroscopy, and second harmonic generation, 
for detecting and tracking topological phase transitions. More broadly, our work motivates further exploration of nonlinear photocurrents as a sensitive probe of other electronic phase transitions, including Lifshitz transitions and interaction-driven topological phases~\cite{Agarwal2024LT2, Agarwal2025LT1, FrankKoppen2025T}.





\textcolor{blue}{\it Acknowledgement.-} We thank Sayan Sarkar for  Fig.~\ref{Figure1}(c). We acknowledge stimulating discussions with Prof. Bahadur Singh,  Prof. Barun Ghosh, Dr. Kamal Das, and Dr. Atasi Chakraborty. A. A. acknowledges funding from the Core Research Grant by Anusandhan National Research Foundation (ANRF, Sanction No. CRG/2023/007003), Department of Science and Technology, India.  We acknowledge the high-performance computing facility at IIT Kanpur, including HPC 2013, and Param Sanganak. R. A. acknowledges funding from IIT Kanpur as an institute postdoctoral fellow. D. D. acknowledges the support from Nanyang Technological University and IIT Kanpur, where the work has been initiated. 

\bibliography{ref.bib}
\vspace{0.3 cm}
\vspace{0.3 cm}
{\bf ENDMATTER}
\vspace{0.3 cm} \\
\textcolor{blue}{\it EM1: ${\cal Z}_2$ indices and invariants on 2D planes--} 
In three-dimensional time-reversal symmetric insulators, topological phases are characterized by a set of four ${\cal Z}_2$ indices, denoted as ${\cal Z}_2 \equiv (\nu_0; \nu_1, \nu_2, \nu_3)$. These indices are constructed from two-dimensional ${\cal Z}_2$ invariants, $\Delta$, evaluated on time-reversal-invariant momentum (TRIM) planes in the three-dimensional Brillouin zone (BZ)~\cite{Z2pack_PRB_2017, PhysRevB.83.035108, Bernevig2011_Z2}. Specifically,
\begin{align}
    \nu_0 &= \sum_{i=1}^{3}\left[ \Delta(k_i = 0) + \Delta(k_i = 0.5) \right] \mod 2~, 
    \label{STI_invariant} \\
    \nu_i &= \Delta(k_i = 0.5)~, \quad i = 1, 2, 3~.
    \label{WTI_invariant}
\end{align}
Here, $k_i$ are dimensionless crystal momenta defined in terms of reciprocal lattice vectors $\vec{g}_i$, such that $\vec{k} = k_1 \vec{g}_1 + k_2 \vec{g}_2 + k_3 \vec{g}_3$. The function $\Delta(k_i = 0)$ or $\Delta(k_i = 0.5)$ denotes the 2D topological invariant calculated on the six time reversal invariant planes of the 3D BZ, where each $\vec{k}$ satisfies $\vec{k} = -\vec{k}$ up to a reciprocal lattice vector.

These six momentum planes can each be regarded as effective 2D time-reversal planes for symmetric insulators. We compute their ${\cal Z}_2$ invariants by tracking the evolution of hybrid Wannier charge centers using the \texttt{Z2Pack} software package~\cite{Z2pack_PRB_2017}. For further methodological details, see Sec.~S2 of the Supplemental Material.

\begin{figure}[t!]
    \centering    
    \includegraphics[width=1.0\linewidth]{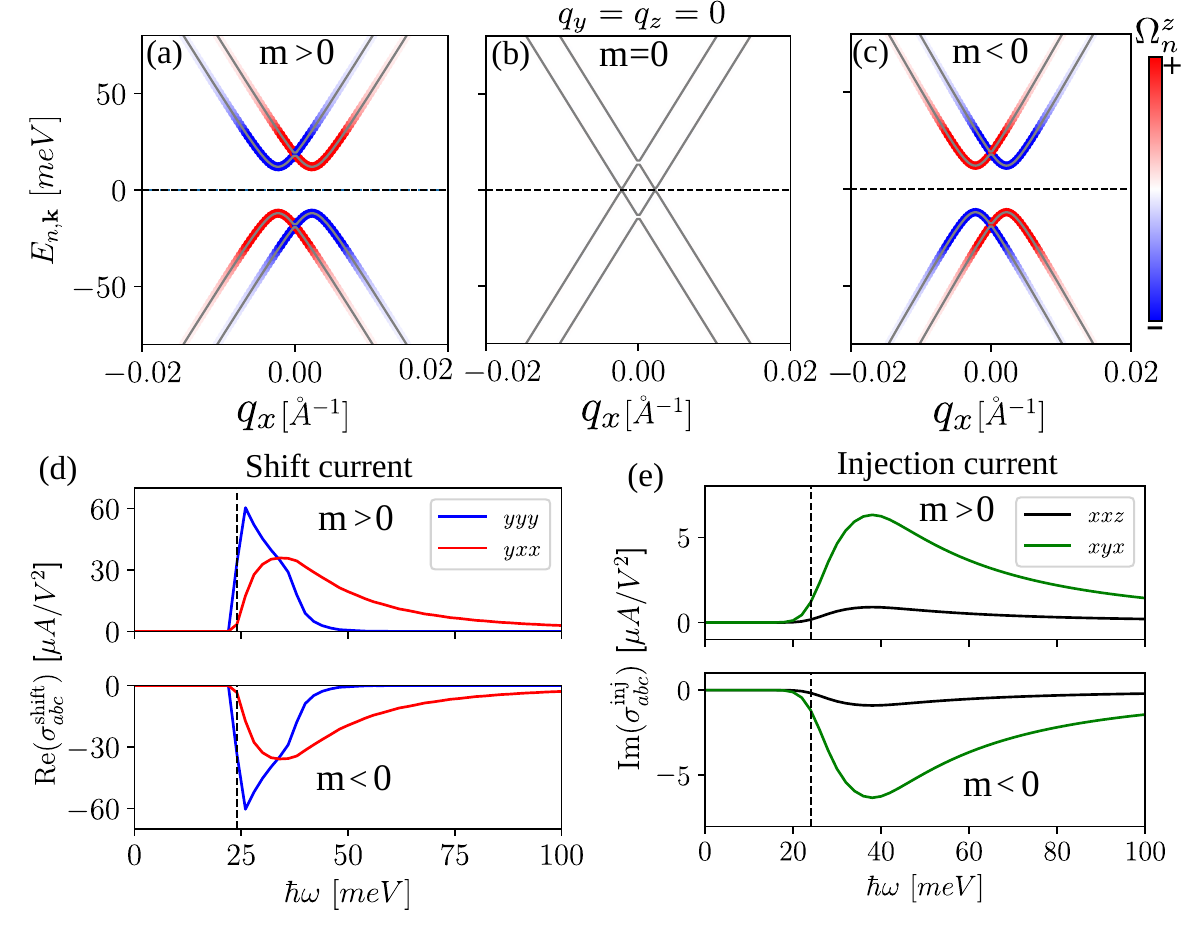}
    \caption{\textbf{Sign reversal of photocurrents with $m$ from a low energy model}. (a-c) Band dispersion of a massive Dirac model with broken inversion symmetry [Eq.~\eqref{bulk_Hk}] with $m$ parameter. At $m=0$, the gap-closing occurs away from TRIM $\vec{q}=0$. The colormap shows z-component of the Berry curvature, which changes sign for $m>0$ and $m<0$, respectively. (d-e) Calculated  shift, $\sigma_{abc}^{\rm shift}$ and injection  $\sigma_{abc}^{\rm inj}$ photo-conductivities with photon energy ($\hbar\omega$) with $m>0$ and $m<0$, respectively. This shows that all these conductivity tensors flip sign for $m>0$, and $m<0$, respectively. We use $|m|=12$ meV, $\delta=0.8$ meV for numerical calculations. 
    }
    \label{Figure_low_energy_model}
\end{figure}

\begin{figure}[t!]
    \centering    
    \includegraphics[width=0.9\linewidth]{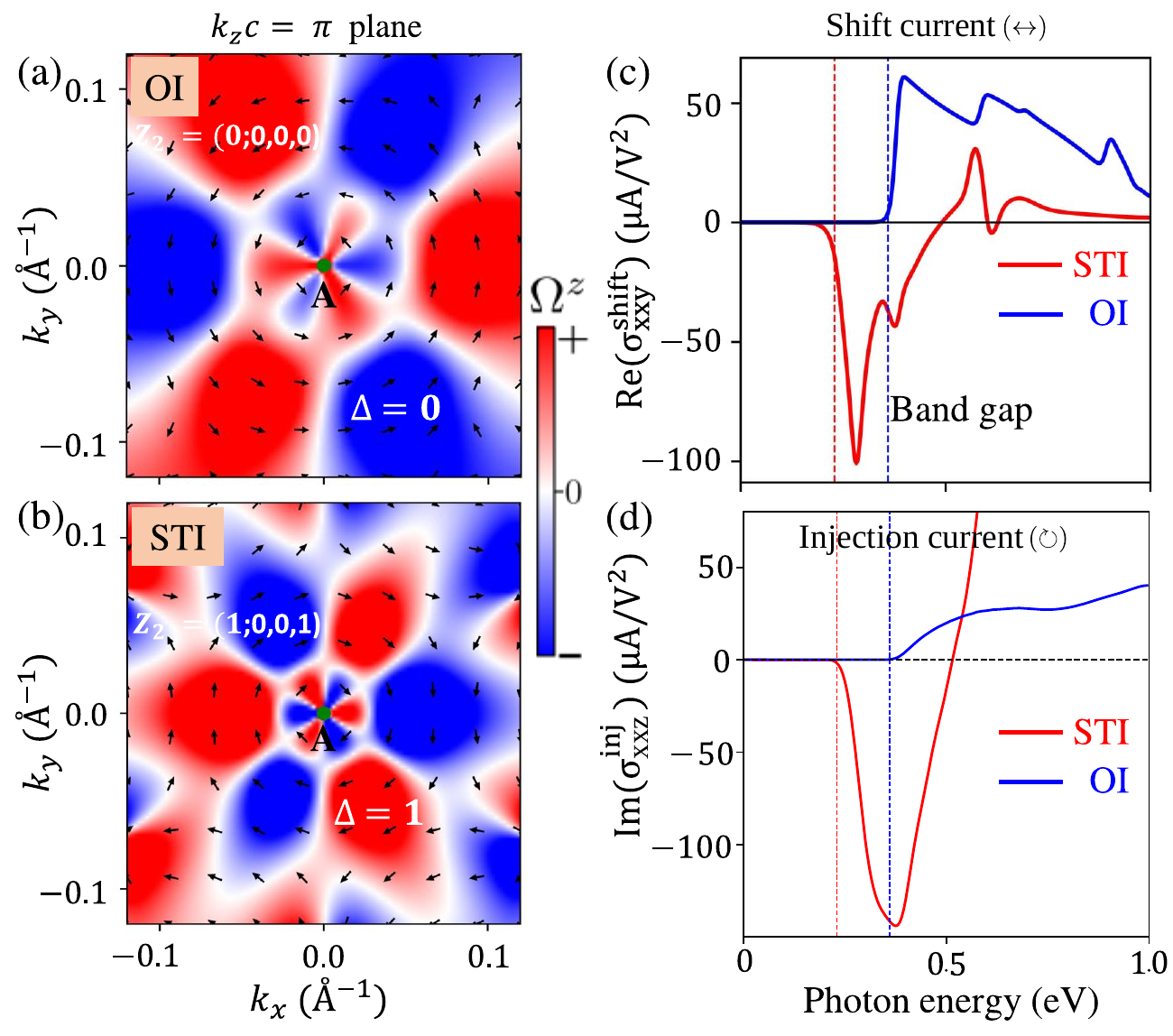}
    \caption{ \textbf{Sign reversal of photocurrents in bulk BiTeI across OI to STI phase transition}.
    (a-b) Berry curvature distribution for top valence band in the OI and STI phases at $k_z=\pi/c$ plane TRIM plane. A clear sign reversal is observed around the A point due to band inversion. It is also consistent with a change in the 2D ${\cal Z}_2$ invariant, $\Delta$ at that plane. (c–d) Calculated shift and injection photocurrents with photon energy in two phases. The vertical lines indicate the bandgap in the two phases, for the chosen parameters for each phase. Distinct sign changes are observed between OI and STI phases near peak values around the band-gap, arising from photoexcitation around the A point. At the higher frequency, the sign change is not evident due to the photo-excitation between the bands not exhibiting band inversion.
    }
    \label{Fig_BiTeI}
\end{figure} %

\textcolor{blue}{{\it EM2:} Photocurrent reversal in $\vec{k} \cdot \vec{p}$ model--}
The ${\cal Z}_2$ topological phase transition in nonmagnetic systems is typically associated with a band inversion mechanism, wherein the bulk band gap closes and reopens around a TRIM point in the Brillouin zone. To capture this behavior, we consider a minimal $\vec{k} \cdot \vec{p}$ Hamiltonian expanded around a specific TRIM point $\vec{k}_0$, in the basis $|\Psi_{\vec{k}}\rangle = \{|\Psi_{+}^{\uparrow}\rangle, |\Psi_{-}^{\uparrow}\rangle, |\Psi_{+}^{\downarrow}\rangle, |\Psi_{-}^{\downarrow}\rangle\}$~\cite{PhysRevLett.115.176404, Ando2022}. For example, a low-energy model for ZrTe$_5$ near the $\Gamma$ point, valid up to linear order in momentum, is given by ~\cite{PhysRevLett.115.176404, Ando2022}, 
\begin{eqnarray}
H(\vec{q},m) &=& \hbar\bigl(v_x q_x\,\sigma_3\otimes\tau_1 
  + v_yq_y\,\sigma_1\otimes\tau_1 
  + v_zq_z\,\mathbb{1}\otimes\tau_2\bigr) \nonumber \\
& + & m\,(\mathbb{1}\otimes\tau_3) +\delta\,(\mathbb{1}\otimes\tau_1 
  + \sigma_1\otimes\tau_2)~.
\label{bulk_Hk}
\end{eqnarray}
Here, $\vec{q} = \vec{k} - \vec{k}_0$ is the wavevector measured from the TRIM point. The mass parameter $m$ controls the bulk band gap inversion and governs the topological character.
The Pauli matrices $\sigma_i$ and $\tau_j$ act on spin and orbital (parity) degrees of freedom, respectively. The term $\delta$ breaks inversion symmetry and arises from slight displacements of Te$_z$ atoms. The velocity components are given by $v_x = 9 \times 10^5$ m/s, $v_y = 0.3 \times 10^5$ m/s, and $v_z = 1.9 \times 10^5$ m/s~\cite{PhysRevB.104.125439}.

\begin{table*}[t]
\centering
\caption{Summary of topological transitions and nonlinear photocurrent signatures across different materials studied in this work.}
\begin{tabular}{ccccc}
\hline
\hline
\textbf{~Material~} & \textbf{~Phase Transition~} & \textbf{~TRIM Plane~} & \textbf{~Topological Change~} & \textbf{~Photocurrent Sign Flip~} \\
\hline
Bi$_2$Te$_3$ & STI $\rightarrow$ WTI & $k_z = \pi/c$ & $\nu_3: 0 \rightarrow 1$ & Shift, Injection \\
\hline
ZrTe$_5$ & STI $\rightarrow$ WTI & $k_y = 0$ & $\Delta(k_y=0): 0 \rightarrow 1$ & Shift, Injection \\
\hline
BiTeI & OI $\rightarrow$ STI & $k_z = \pi/c$ & $\nu_3: 0 \rightarrow 1$ & Shift, Injection \\
\hline
\hline
\end{tabular}
\label{tab:summary}
\end{table*}

 By setting the inversion-symmetry-breaking term $\delta = 0$ in Eq.~\eqref{bulk_Hk}, the Hamiltonian simplifies to a form that preserves inversion symmetry, facilitating analytical analysis. The resulting low-energy Hamiltonian near the $\Gamma$ point is, 
\begin{equation}
H  =
\begin{pmatrix}
 m & q_\parallel e^{-i\phi} & 0 & q_\perp \\
 q_\parallel e^{i\phi} & -m & q_\perp & 0 \\
 0 & q_\perp & m & -q_\parallel e^{i\phi} \\
 q_\perp & 0 & -q_\parallel e^{-i\phi} & -m
\end{pmatrix}~,
\label{eq:H}
\end{equation}%
where $q_\parallel e^{i\phi} = \hbar (v_x q_x+iv_z q_z)$,  $q_\parallel=\hbar \sqrt{v_x^2 q_x^2+v_z^2 q_z^2}$ and $q_\perp=\hbar v_y q_y$.
The energy eigenvalues of this Hamiltonian are 
$E^{s_1 s_2} = s_1 \sqrt{m^2 + q_\parallel^2 + q_\perp^2}$,
where $s_i$ are signs $\pm$ that label the four bands. 
In the $q_\perp = 0$ plane (e.g., $k_y=0$), the Berry curvature for the $y$-component is
\begin{equation}
\Omega_y^{s_1 s_2} = -s_2 \frac{m}{(m^2 + q_\parallel^2)^{3/2}}~.
\end{equation}
The mass parameter $m$ governs the topological phase transition between the weak topological insulator (WTI, $m>0$) and strong topological insulator (STI, $m<0$) phases. As $m$ changes sign, a band inversion occurs at the $\Gamma$ point, flipping the parity of the occupied bands along with the $\mathcal{Z}_2$ topological invariant associated with the parity eigenvalues. 
The sign change of $m$ also reverses the Berry curvature and the shift vector for a given band in the TRIM plane, which reflects in the polarity reversal of the injection and shift currents.


At the critical point ($m = 0$), the gap closes at a finite momentum $\vec{q} \neq 0$, due to the presence of the inversion breaking term $\delta$. The resulting electronic structure along with the Berry curvature is shown in Fig.~\ref{Figure_low_energy_model}(a - c). The band inversion leads to a reversal in the distribution of Berry curvature, which underlies the sign change in nonlinear photocurrents. Similar redistribution is also seen in the shift vector across the transition.  

 The resulting shift and injection conductivities for opposite values of the mass parameter ($m = \pm 12$ meV), are shown in Fig.~\ref{Figure_low_energy_model}(d–e). Both the $\sigma_{yxx}^{\rm shift}$ and $\sigma_{xxz}^{\rm inj}$ conductivities reverse sign across the topological transition. These calculations clearly establish that the polarity of nonlinear optical responses is governed by the mass term $m$ associated with the band inversion and the topological phase transition. 

\textcolor{blue}{{\it EM3: Reversal across OI to STI transition in \text{BiTeI}--}}
Both the cases studied in the main manuscript, Bi$_2$Te$_3$ and ZrTe$_5$ show sign reversal in photocurrents across an WTI to STI transition. Here, we briefly demonstrate the same behavior in a different material system, bulk BiTeI (space group- $P3m1$), where sign reversal occurs across the OI to STI phase transition under hydrostatic pressure~\cite{Bahramy2012}. At it's equilibrium volume ($V_0$), BiTeI is in the OI phase. Under compression ($V=0.86V_0$), it transitions into STI phase having ${\cal Z}_2=(1;0,0,1)$, driven by band inversion at $k_z=\pi/c$ TRIM plane around A point [see Fig.~\ref{Figure1}(b)]~\cite{Bahramy2012}. In Fig.~\ref{Fig_BiTeI}(a-b), we show that the Berry curvature reorganizes across the transition near A point, which leads to a sign reversal of bulk photocurrents [Fig.~\ref{Fig_BiTeI}(c-d)]. The $P3m1$ space group admits four nonvanishing elements (i.e., $xxz=yyx$, $xxy=yxx=-yyy$, $zxx=zyy$, and $zzz$) of shift conductivity and one independent component ($xxz=-xzx=yyz=-yzy$) of the injection conductivity tensor~\cite{gallego2019automatic}. We only show $\sigma^{\rm shift}_{xxy}$ and $\sigma^{\rm inj}_{xxz}$ components in Fig.~\ref{Fig_BiTeI}(c-d). The other components of shift conductivities are shown in Fig.~S7 of the SM~\cite{Note1}. 

Finally, we summarize all the studied topological transitions and associated photocurrent responses in Table~\ref{tab:summary}. 
\end{document}